\documentclass[vecarrow]{svmult}

\usepackage{makeidx}         
\usepackage{graphicx}        
\usepackage{psfrag}          
\usepackage{multicol}        
\usepackage{amsmath}
\usepackage{amssymb}
\usepackage{amsfonts}
\usepackage{latexsym}
\usepackage[T1]{fontenc}
\usepackage{subfigure}                      
\usepackage{float}
\usepackage{mathrsfs}
\usepackage{color}

\begin{document}

\title{On the Influence of Stochastic Moments in the Solution of the Neutron Point Kinetics Equation}

\titlerunning{On the Influence of Stochastic Moments}

\author{M. Wollmann da Silva \and B.E.J. Bodmann \and M.T. Vilhena \and R. Vasques}

\authorrunning{M. Wollmann da Silva et al.}

\institute{Universidade Federal do Rio Grande do Sul, Porto Alegre, Brazil;
\texttt{milena.wollmann@ufrgs.br}, \texttt{bardobodmann@ufrgs.br},\\
\texttt{mtmbvilhena@gmail.com}, \texttt{richard.vasques@fulbrightmail.org}}

\maketitle

\newcommand{\oH}{{\mathaccent'27 H}}

\section{Introduction}

The neutron point kinetics equations, which model the time-dependent behaviour of nuclear reactors \cite{abha03,haal05,he71,sa89}, are often used to understand the dynamics of nuclear reactor operations (e.g. power fluctuations caused by control rod motions during start-up and shut-down procedures). They consist of a system of coupled differential equations that model the interaction between (i) the neutron population, and (ii) the concentration of the delayed neutron precursors, which are radioactive isotopes formed in the fission process that decay through neutron emission. These equations are deterministic in nature, and therefore can provide only average values of the modeled populations. However, the actual dynamical process is stochastic: the neutron density and the delayed neutron precursor concentrations vary randomly with time.

To address this stochastic behaviour, Hayes and Allen \cite{haal05} have generalized the standard deterministic point kinetics equations. They have derived a system of stochastic differential equations that can accurately model the random behaviour of the neutron density and the precursor concentrations in a point reactor. Due to the issue of stiffness, they numerically implement this system using a stochastic piecewise constant approximation method (Stochastic PCA).

Here, we present a study on the influence of stochastic fluctuations upon the results of the neutron point kinetics equations. We reproduce the stochastic formulation introduced in \cite{haal05} and compute Monte Carlo numerical results for examples with constant and time-dependent reactivity,
comparing these results  with stochastic and deterministic methods found in the literature \cite{haal05,ra12,sile14}.

The remainder of this work is organized as follows. In Section \ref{sec2} we reproduce the derivation of the stochastic equations introduced in \cite{haal05}. Section \ref{sec3} starts with a short discussion on the numerical implementation of the stochastic models. In Section \ref{sec3.1} we provide numerical results for examples with constant reactivity, for the cases of one and six precursor groups; and in Section \ref{sec3.2} we present results for an example with linear reactivity and one precursor group. Finally, in Section \ref{sec4} we discuss the stochastic fluctuations we have encountered, and address the future steps to be undertaken in order to accurately study them.

\section{Stochastic Model Formulation}\label{sec2}

In this section, we reproduce the stochastic formulation introduced by Hayes and Allen. Following \cite{haal05,he71}, the time-dependent equations that describe the neutron density and the delayed neutron precursor concentrations are given by
\begin{subequations}\label{eq1}
\begin{align}
&\frac {\partial N} {\partial t} = D v \nabla^2 N - \left ( \Sigma_a - \Sigma_f \right ) v N + \left [ \left ( 1 - \beta \right ) k_ {\infty} \Sigma_a - \Sigma_f \right ] v N + \sum_i \lambda_i C_i + S_0,& 
\label{eq1a} \\
&\frac {\partial C_i} {\partial t} = \beta_i k_{\infty} \Sigma_a v N - \lambda_i C_i,&
\label{eq1b}
\end{align}
\end{subequations}
where $i=1,2,...,m$, $v$ is the velocity, $N=N(r,t)$ is the neutron density at position $r$ and time $t$, and $C_i=C_i(r,t)$ is the concentration of the $i$-th type of precursor at position $r$ and time $t$. In the right-hand side of Eq.\ (\ref{eq1a}) we have the following terms:
\begin{itemize}
\item $Dv\nabla^2 N$, representing the diffusion of neutrons.
\item $(\Sigma_a - \Sigma_f)vN$, representing the capture of neutrons. Notice that the capture cross-section is given by the difference between the absorption ($\Sigma_a$) and the fission ($\Sigma_f$) cross-sections.
\item $\left[ \left( 1-\beta\right)k_ {\infty} \Sigma_a-\Sigma_f\right]vN$, representing the prompt-neutron contribution to the source. Here, $\displaystyle{\beta=\sum_{i=1}^{m} \beta_i}$ is the delayed-neutron fraction and $k_\infty$ is the infinite medium reproduction factor.
\item $\sum_{i}\lambda_i C_i$, representing the rate of transformation from the neutron precursors to the neutron population, with $\lambda_i$ as the decay constant.
\item $S_0(r,t)$, representing the external source.
\end{itemize}
 
Assuming that $N$ and $C_i$ are separable in time and space, we can write $N(r,t) = f(r) n(t)$ and $C_i(r, t) = g_i(r) c_i(t)$, where $n(t)$ and $c_i(t)$ represent the total neutron density and the total concentration of precursors of the $i$-th type at time $t$, respectively. Equations (\ref{eq1}) now become
\begin{subequations}
\begin{align*}
\frac{dn}{dt}(t)&=Dv\frac{\nabla^2f(r)}{f(r)}n(t)-(\Sigma_a - \Sigma_f) vn(t) \\
 &\hspace{2 cm} +[( 1-\beta)k_{\infty}
  \Sigma_a - \Sigma_f]vn(t) + \sum_{i}\lambda_i \frac{g_i(r)c_i(t)}{f(r)} + \frac{S_0(r,t)}{f(r)},
  \\[5pt]
 \frac{dc_i}{dt}(t)&=\beta_i k_{\infty}\Sigma_a v \frac{f(r)n(t)}{g_i(r)}-\lambda_i c_i(t).
\end{align*}
\end{subequations}
\setcounter{equation}{1}
It is assumed that (i) $\frac{f(r)}{g_i(r)}=1$; (ii) $f$ satisfies $\nabla^2f+B^2f=0$; and (iii) $S_0$ has the same spatial dependence as $f$. If we write $q(t)=\frac{S_0(r,t)}{f(r)}$, the previous equations become
\begin{subequations}\label{eq2}
\begin{align}
&\frac {dn} {dt} = -D v B^2 n - (\Sigma_a - \Sigma_f) v n + [ ( 1 - \beta ) k_{\infty} \Sigma_a - \Sigma_f ] v n + \sum_{i} \lambda_i c_i + q,
\label{eq2a} \\
&\frac {dc_i} {dt} = \beta_i k_{\infty} \Sigma_a v n -\lambda_i c_i.&
\label{eq2b}
\end{align}
\end{subequations}
Furthermore, the terms in Eq.\ (\ref{eq2a}) can be rearranged according to the type of neutron reaction:
\begin{equation}\label{eq3}
\frac {dn} {dt} = \underbrace{-D v B^2 n - (\Sigma_a - \Sigma_f) v n}_{deaths} + \underbrace{( k_{\infty} \Sigma_a - \Sigma_f ) v n}_{births} - \underbrace{\beta k_{\infty} \Sigma_a v n + \sum_{i} \lambda_i c_i}_{transformations} + q.
\end{equation}

In order to simplify the notation, several parameters are now introduced. We define the absorption lifetime $l_{\infty}=\frac{1}{v\Sigma_a}$ and the diffusion length $L^2= \frac{D}{\Sigma_a}$, and rewrite Eqs.\ (\ref{eq3}) and (\ref{eq2b}) as
\begin{subequations}\label{eq4}
\begin{align}
&\frac {dn} {dt} = \underbrace{ \left [ \frac {-L^2 B^2 - \frac {(\Sigma_a - \Sigma_f)} {\Sigma_a}} {l_{\infty}} \right ] }_{deaths} n + \underbrace{ \left [ \frac {k_{\infty} - \frac {\Sigma_f} {\Sigma_a}} {l_{\infty}} \right ] }_{births} n - \underbrace{ \frac {\beta k_{\infty}} {l_{\infty}} n + \sum_i \lambda_i c_i}_{transformations} + q,& 
\\
&\frac {d c_i} {dt} = \frac {\beta_i k_{\infty}} {l_{\infty}} n - \lambda_i c_i.&
\end{align}
\end{subequations}
Defining the reproduction factor $k = \frac {k_{\infty}} {1 + L^2 B^2}$ and the neutron lifetime $l_0 = \frac {l_{\infty}} {1 + L^2 B^2}$, Eqs.\ (\ref{eq4}) become
\begin{subequations}\label{eq5}
\begin{align}
  \frac{dn}{dt}&=\left[-\frac{1}{l_0}+\frac{\Sigma_f}{\Sigma_a l_{\infty}} \right]n+\left[\frac{k}{l_0}-\frac{\Sigma_f}{\Sigma_a l_{\infty}} \right]n
  -\frac{\beta k}{l_0}n+\sum_i \lambda_ic_i +q, 
\\
\frac{dc_i}{dt}&=\frac{\beta_i k}{l_0}n-\lambda_ic_i. 
\end{align}
\end{subequations}
Next, we introduce the neutron generation time $l=\frac{l_0}{k}$. Substituting $l$ into Eqs.\ (\ref{eq5}), we obtain
\begin{subequations}
\begin{align*}
  \frac{dn}{dt}&=\left[-\frac{1}{kl}+\frac{\Sigma_f}{\Sigma_a l_{\infty}} \right]n+\left[\frac{1}{l}-\frac{\Sigma_f}{\Sigma_a l_{\infty}} \right]n
  -\frac{\beta}{l}n+\sum_i \lambda_ic_i +q,
\\[5pt]  \frac{dc_i}{dt}&=\frac{\beta_i}{l}n-\lambda_ic_i.
\end{align*}
 \end{subequations}\setcounter{equation}{5}
Finally, we define reactivity $\rho=1-\frac{1}{k}$. Moreover, a simple algebraic calculation shows that $\frac{\Sigma_f}{\Sigma_a l_{\infty}} = \frac{\alpha}{l}$, where $\alpha=\frac{\Sigma_f}{\Sigma_a  k_{\infty}}\approx\frac{1}{\nu}$ and $\nu$ is the number of neutrons per fission. Hence, the final deterministic system becomes
\begin{subequations}\label{eq6}
\begin{align}
 \frac{dn}{dt}&=\underbrace{-\left[\frac{-\rho+1-\alpha}{l}\right]}_{deaths}n+
 \underbrace{\left[\frac{1-\alpha-\beta}{l}\right]}_{births}n+\underbrace{\sum_i \lambda_ic_i}_{transformations} +q, 
\\[5pt] \frac{dc_i}{dt}&=\frac{\beta_i}{l}n-\lambda_ic_i,
\end{align}
\end{subequations}
for $i=1,2...,m$.

To derive the stochastic system, we first consider the case of just one precursor; that is, $\beta=\beta_1$. (The system will be generalized to $m$ precursors later.) Equations (\ref{eq6}) for one precursor are written as
 \begin{subequations}
\begin{align*}
 \frac{dn}{dt}(t)&=\left\{-\left[\frac{- \rho+1-\alpha}{l} \right]+\left[\frac{1-\alpha-\beta}{l}\right]\right\}n(t) +\lambda_{1}c_{1}(t) + q,
\\[5pt] \frac{dc_{1}}{dt}(t)&=\frac{\beta_{1}}{l}n(t)-\lambda_{1}c_{1}(t).
\end{align*}
\end{subequations}
We consider a time interval $\Delta t$ small enough to guarantee that the probability of more than one event occurring during $\Delta t$ is negligible. Let $[\Delta n, \Delta c_1]^T$ be a random vector variable that represents the changes in the neutron density and in the delayed neutron precursor concentration. The four possible events are
\begin{subequations}
\begin{align*}
 \left[ \begin{array}{c}
      \Delta n\\      \Delta c_1\\
     \end{array}\right]_1&=\left[ \begin{array}{c}
                                  -1\\ 0\\
                                 \end{array}
\right] = \text{death (capture),}\\[5pt]
 \left[ \begin{array}{c}
      \Delta n\\      \Delta c_1\\
     \end{array}\right]_2&=\left[ \begin{array}{c}
                                  -1+(1-\beta)\nu\\ \beta_1 \nu\\
                                 \end{array}
\right]=\begin{array}{c} \text{birth (fission event and}\\\text{ production of delayed neutrons),}\end{array}\\[5pt]
 \left[ \begin{array}{c}
      \Delta n\\      \Delta c_1\\
     \end{array}\right]_3&=\left[ \begin{array}{c}
                                  1\\ -1\\
                                 \end{array}
\right]=\begin{array}{c} \text{transformation of a delayed}\\\text{neutron precursor to a neutron,}\end{array}\\[5pt]
 \left[ \begin{array}{c}
      \Delta n\\      \Delta c_1\\
     \end{array}\right]_4&=\left[ \begin{array}{c}
                                  1\\ 0\\
                                 \end{array}
\right]= \text{birth of a source neutron};
\end{align*}
\end{subequations}
and the probabilities of these events (assuming $\alpha= \frac{1}{\nu}$) are:
\begin{subequations}
\begin{align*}
P_1 = \left(\frac{-\rho+1-\alpha}{l}\right) n\Delta t, \quad P_2 = \left(\frac{1}{\nu l}\right)n\Delta t, \quad P_3 = \lambda_1 c_1 \Delta t,\quad P_4 =q \Delta t.
\end{align*}
\end{subequations}
It is also assumed that the neutron source produces neutrons randomly following a Poisson process with intensity $q$.

Finally, the mean change $E([\Delta n, \Delta c_1]^T)$ for the small time interval $\Delta t$ is given by
\begin{align*}\label{mean}
 E\left(\left[ \begin{array}{c}
      \Delta n\\      \Delta c_1\\
     \end{array}\right]\right)=\sum_{k=1}^{4}P_k\left[ \begin{array}{c}
      \Delta n\\      \Delta c_1\\
     \end{array}\right]_k=\left[ \begin{array}{c}
                                  \frac{p-\beta}{l}n+\lambda_1c_1+q\\ \frac{\beta_1}{l}n-\lambda_1c_1
                                 \end{array}
\right]\Delta t,
\end{align*}
and the covariance of the change is given by
\begin{subequations}
\begin{align*}
E\left(\left[ \begin{array}{c}
      \Delta n\\      \Delta c_1\\
     \end{array}\right]\left[\begin{array}{cc}
                        \Delta n       \Delta c_1\\
                       \end{array}\right]\right)&=\sum_{k-1}^{4}P_k\left[ \begin{array}{c}
      \Delta n\\      \Delta c_1\\
     \end{array}\right]\left[\begin{array}{cc}
                        \Delta n       \Delta c_1\\
                       \end{array}\right]_k=\hat{B}\Delta t,
\end{align*}
where $\hat{B}$ is defined as
\begin{align*}
 \hat{B}&=\left[ \begin{array}{cc}
         \gamma n+\lambda_1c_1 + q &\frac{\beta_1}{l}(-1+(1-\beta)\nu)n -\lambda_1c_1 \\
         \frac{\beta_1}{l}(-1+(1-\beta)\nu)n -\lambda_1c_1 & \frac{\beta_1^2 \nu}{l}n+\lambda_1c_1\\
        \end{array}
\right]
\end{align*}
\end{subequations}
and $\gamma=\frac{-1-\rho+2\beta+(1-\beta)^2\nu}{l}$.

With the assumption that the changes are approximately normally distributed, the above results imply that, to $O((\Delta t)^2)$,
\begin{equation*}
\left[ \begin{array}{c}
n(t+\Delta{t}) \\
c_1(t+\Delta{t}) \\
\end{array} \right ] =
\left[ \begin{array}{c}
n(t)\\
c_1(t)\\
\end{array} \right ] + \hat{A} \left [ \begin{array}{c}
n(t) \\
c_1(t) \\
\end{array} \right ] \Delta t +
\left[
\begin{array}{c} 
q \\ 0 \\
\end{array} \right ] \Delta t
+ \hat{B}^\frac{1}{2} \sqrt{\Delta{t}} \left [ \begin{array}{c}
\eta_1 \\ 
\eta_2
\end{array} \right ],
\end{equation*}
where $ \eta_1, \eta_2 \sim \mathcal{N}(0,1)$, $\hat{B}=\hat{B}^\frac{1}{2}\cdot\hat{B}^\frac{1}{2}$, and $\hat{A} = \left [ \begin{array}{c} \frac {p - \beta} {l} + \lambda_1 \\ \frac {\beta_1} {l} - \lambda_1 \end{array} \right ]$.

As $\Delta t\rightarrow0$, the above equations yield the following It\^o stochastic differential equation system \cite{hi01,rapa13}:
\setcounter{equation}{6} 
\begin{equation}\label{eq7}
\frac {d} {dt} \left [ \begin{array}{c} n \\ c_1 \\ \end{array} \right ] = \hat{A} \left [ \begin{array}{c} n \\ c_1 \\ \end{array} \right ] + \left [ \begin{array}{c} q \\ 0 \\ \end{array} \right ] + \hat{B}^\frac{1}{2} \frac {d \vec{W}} {dt}, \qquad \vec{W} = \left [ \begin{array}{c} W_1(t) \\ W_2(t) \\ \end{array} \right ],
\end{equation}
where $W_1 (t)$ and $W_2 (t)$ are Wiener processes. Equations (\ref{eq7}) are the stochastic neutron point kinetics equations for one precursor group. 

To generalize these equations to $m$ precursors, let
\begin{equation*}
\hat{A}=\left[\begin{array}{ccccc}
\frac{\rho(t) -\beta}{l} & \lambda_1& \lambda_2&\dots &\lambda_m\\
\frac{\beta_1}{l} & -\lambda_1 & 0 &\dots & 0\\
\frac{\beta_2}{l} & 0 & -\lambda_2 &\dots & 0\\
\vdots & \vdots & \ddots & \ddots & \vdots\\
\frac{\beta_m}{l}&0&\dots&0&-\lambda_m\\
\end{array}\right]
\end{equation*}
and
\begin{equation*}
\hat{B}=\left[\begin{array}{ccccc}
\zeta & a_1& a_2&\dots &a_m\\
a_1 & r_1 & b_{2,3} &\dots & b_{2,m+1}\\
a_2 & b_{3,2} &r_2 &\dots & b_{m,m+1}\\
\vdots & \vdots & \ddots & \ddots & \vdots\\
a_m&b_{m+1,2}&\dots&b_{m+1.m}&r_m\\
\end{array}\right],
\end{equation*}
where
\begin{align*}
\zeta&=\gamma{n}+\sum_{i=1}^{m}\lambda_i c_i +q,
\\[5pt] \gamma&=\frac{-1-\rho+2\beta+(1-\beta)^2\nu}{l},
\\[5pt] a_i&=\frac{\beta_i}{l}(-1+(1-\beta)\nu)n-\lambda_ic_i ,
\\[5pt] b_{i,j}&=\frac{\beta_{i-1}\beta_{j-1}\nu}{l}n,
\\[5pt] r_i&=\frac{{\beta_i}^2 \nu}{l}n+\lambda_ic_i.
\end{align*}
Using the same approach as before, but now for $m$ precursors, we obtain the It\^o stochastic system:
\setcounter{equation}{7}
\begin{equation}\label{eq8}
\frac{d}{dt}\left[ \begin{array}{c}
n(t)\\       c_1(t)\\ c_2(t)\\ \vdots\\ c_m(t)\\
\end{array}\right]= \hat{A}\left[ \begin{array}{c}
n(t)\\       c_1(t)\\ c_2(t)\\ \vdots \\c_m(t)
\end{array}\right]+
\left[\begin{array}{c}
q\\0\\0\\ \vdots \\ 0\end{array}\right]
+\hat{B}^\frac{1}{2}\frac{d\vec{W}}{dt}(t).
\end{equation}
Note that if $\hat B=0$, then Eq.\ (\ref{eq8}) reduces to the standard deterministic point kinetics equations.

\section{Numerical Results}\label{sec3}

We begin this section by briefly sketching the implementation of two approaches that address the stochastic behavior discussed in this work: (I) the Stochastic PCA model \cite{haal05}, and (II) the Euler-Muruyama approximation \cite{ra12}. Specific details of each implementation can be found in the indicated references.
\\

\textbf{(I)} \underline{The Stochastic PCA model} is based on the system given in equation (\ref{eq8}). For instance, assuming $m=6$ delayed groups, this system can be written as
\begin{equation}\label{pca}
 \frac{d
\vec{x}}{dt}=A\vec{x}+B(t)\vec{x}+\vec{F}
(t)+\hat{B}^{\frac{1}{2}}\frac{d\vec{W}}{dt},
\end{equation}
where $\hat{B}$ is already known and
\begin{equation*}
A=\left[
\begin{array}{ccccccc}
\frac{-\beta}{l}&\lambda_1&\lambda_2&\lambda_3&\lambda_4&\lambda_5&\lambda_6\\
\frac{\beta_1}{l}&-\lambda_1&0&0&0&0&0\\
\frac{\beta_2}{l}&0&-\lambda_2&0&0&0&0\\
\frac{\beta_3}{l}&0&0&-\lambda_3&0&0&0\\
\frac{\beta_4}{l}&0&0&0&-\lambda_4&0&0\\
\frac{\beta_5}{l}&0&0&0&0&-\lambda_5&0\\
\frac{\beta_6}{l}&0&0&0&0&0&-\lambda_6\\
\end{array}
\right] ,\,\,\,\,\,
B(t)=\left[
\begin{array}{ccccccc}
\frac{\rho(t)}{l}&0&0&0&0&0&0\\
0&0&0&0&0&0&0\\
0&0&0&0&0&0&0\\
0&0&0&0&0&0&0\\
0&0&0&0&0&0&0\\
0&0&0&0&0&0&0\\
0&0&0&0&0&0&0\\
\end{array}
\right],
\end{equation*}
\[
\vec{F}(t)=\left[ q(t), 0, 0, 0, 0, 0, 0 \right]^T,
\,\,\,\,\,
\vec{x} = \left[n,c_1,c_2,c_3,c_4,c_5,c_6\right]^T.
\]
The source function $q(t)$ and the reactivity function $\rho(t)$ are approximated by piecewise constant functions; in particular
\begin{equation*}
\rho(t)\approx\rho \left(\frac{t_i+t_{i+1}}{2}\right)=\rho_i ,
\,\,\,\,\, \text{for } t_i\leq t\leq t_{i+1}
\end{equation*}
and
\begin{equation*}
B(t)\approx B \left(\frac{t_i+t_{i+1}}{2}\right)=B_i ,
\,\,\,\,\, \text{for } t_i\leq t\leq t_{i+1}.
\end{equation*}
Now, for $t_i\leq t\leq t_{i+1}$, equation (\ref{pca}) becomes
\begin{equation*}
 \frac{d
\vec{x}}{dt}=A\vec{x}+B_i\vec{x}+\vec{F}
(t)+\hat{B}^{\frac{1}{2}}\frac{d\vec{W}}{dt},
\end{equation*}
and using It\^o's formula \cite{kloeden} we obtain
\begin{equation*}
 \frac{d}{dt}\left[ e^{-(A+B_i)t}\vec{x}\right]=
e^{-(A+B_i)t}\vec{F}(t)+
e^{-(A+B_i)t}\hat{B}^{\frac{1}{2}}\frac{d\vec{W}}{dt}.
\end{equation*}
Finally, this equation is approximated using Euler's method, and the eigenvalues and eigenvectors of the matrix $(A+B_i)$ are computed using diagonalization.
\\

\textbf{(II)} \underline{The Euler-Maruyama approximation} performs the time-discrete approximation of an It\^o process. Let $\{X_t\}$ be an It\^o process on $t\in[t_0,T]$ that satisfies the stochastic differential equation $dX_t = a(t,X_t) dt + b(t,X_t) dW_t$, $X_{t_0} = X_0$. For a given time-discretization $t_0<t_1<t_2<...<t_N=T$, an Euler-Maruyama approximation is a continuous time stochastic process $\{Y(t), t_0\leq t\leq T\}$ that satisfies the interactive scheme given by \cite{kloeden} 
\begin{equation*}
Y_{n+1}=Y_n +a(t_n,Y_n)\Delta t_{n+1}+b(t_n,Y_n)\Delta W_{n+1},  \,\,\,\,\, n=0,1,...,N-1,
\end{equation*}
where $Y_0=X_0$, $Y_n=Y(t_n)$, $\Delta t_{n+1}=t_{n+1}-t_n$, and $\Delta W_{n+1}=W(t_{n+1})-W(t_n)$. Each random number is given by $\Delta W_n=z_n \sqrt{\Delta t_n}$, where $z_n$ is chosen from a standard normal distribution $\mathcal{N}(0,1)$. In this type of procedure the considered time intervals must be equidistant.
\\

In the following sections we consider examples with constant and linear reactivity, and present Monte Carlo (MC) simulations for each one of them. We compare the MC estimates to the results obtained with the stochastic models previously discussed, as well as with the Deterministic Model \cite{pevi09, sile14}.

In these MC simulations, we have chosen the time interval $\Delta t$ to be small enough such that the likelihood of more than one event taking place during $\Delta t$ is very small. This was achieved by considering the half-life time of the precursor groups, according to the time decay constants $\lambda_i$. 
The number of seeds used in the MC estimates for each case was large enough to guarantee that the statistical error of the mean values is less than 0.05\% (with 95\% confidence).

\subsection{Constant Reactivity}\label{sec3.1}

In the following examples, we present the results for the mean values $E$ of the neutron density and the delayed neutron precursors concentration. In addition, we also present the standard deviations $\sigma$ of these quantities for the stochastic models. 
\begin{table}[!ht]
\center
\caption{Results for one precursor group and reactivity $\rho = -1/3$.}
\begin{tabular}{||c|c|c|c|c||} \hline\hline
 &  Monte & Stochastic &Euler-Maruyama&  Deterministic  \\ 
 &  Carlo     & PCA & approximation & model \\ \hline
 ${E(n(2))}$ &400.032 & 395.32 &412.23& 412.13\\
${\sigma(n(2))}$ &27.311 & 29.411&34.391& --\\ \hline
${E(c_1(2))}$ & 300.01 & 300.67 &315.96& 315.93\\
${\sigma(c_1 (2))}$ & 7.807 & 8.3564 &8.2656& -- \\ \hline\hline
\end{tabular}\label{tab1}
\end{table}
 
In the first example, we reproduce a test case presented in \cite{haal05}, which assumes only one neutron precursor and simulates a step-reactivity insertion. Although it does not model an actual physical nuclear reactor problem, it provides simple computational solutions for comparison with Monte Carlo results. The parameters are $\lambda_1=0.1$, $\beta_1=0.005$, $\nu=2.5$, $q=200$, $l=\frac{2}{3}$, and $\rho=-\frac{1}{3}$, with equilibrium values for the initial condition: $\vec{x}(0)=[400,300]^T$.

Table \ref{tab1} shows that, while the standard deviations for both quantities are one order of magnitude smaller than their correspondent mean values, the standard deviation for the neutron density is still significant ($\approx 7\%$ of the mean). This suggests that a deterministic approach may not be sufficient for the computation of this quantity.
\begin{table}[!ht]
\center
\caption{Results for six precursor groups and reactivity $\rho=0.003$.}
\begin{tabular}{||c|c|c|c|c||} \hline\hline
 &  Monte & Stochastic &Euler-Maruyama&  Deterministic  \\ 
 &  Carlo           &  PCA &approximation& model \\ \hline
$E(n(0.1))$ &183.04 & 186.31 &208.6& 200.005\\
$\sigma(n(0.1))$ &168.79 &164.16&255.95& --\\ \hline
$E(\sum_{i=1}^{6}c_i(0.1))$ & 4.478$\times 10^5$ &  4.491$\times 10^5$ &$4.498 \times 10^5$& 4.497$\times 10^5$ \\
$\sigma(\sum_{i=1}^{6}c_i(0.1))$ &1495.72 & 1917.2 &1233.38&-- \\ \hline\hline
\end{tabular}\label{tab2}
\end{table}

The next example (two scenarios) uses $m=6$ delayed neutron precursor groups, and models step reactivity insertions for an actual nuclear reactor \cite{chat85,haal05,kial04}. The first scenario models a prompt insertion with $\rho=0.003$, whereas the second scenario models a prompt insertion with $\rho = 0.007$. In both scenarios the parameters are chosen as follows:
\begin{align*}
\lambda_i&=[0.0127,0.0317,0.115,0.311,1.4,3.87];\\ \beta_i&=[0.000266,0.001491,0.001316,0.002849,0.000896,
0.000182];\\
\beta &= 0.007;\,\,\,\,\nu=2.5;\,\,\,\, q=0; \,\,\,\, l=0.00002;
\end{align*}
with an initial condition that assumes a source-free equilibrium:
\begin{equation*}
 \vec{x}(0)=100\left[
                          \frac{\beta_1}{\lambda_1 l},
                          \frac{\beta_2}{\lambda_2 l},
                          \frac{\beta_3}{\lambda_3 l},
                          \frac{\beta_4}{\lambda_4 l},
                          \frac{\beta_5}{\lambda_5 l},
                          \frac{\beta_6}{\lambda_6 l}
                          \right]^T.
\end{equation*}
\begin{table}[!ht]
\center
\caption{Results for six precursor groups and reactivity $\rho=0.007$.}
\begin{tabular}{||c|c|c|c|c||} \hline\hline
 &  Monte & Stochastic  &Euler-Maruyama&  Deterministic  \\ 
 &  Carlo    &  PCA &approximation& model \\ \hline
$E(n(0.001))$ &135.66 & 134.55 &139.568& 139,61\\
$\sigma(n(0.001))$ &93.376 &91.242&92.042& --\\ \hline
$E(\sum_{i=1}^{6}c_i(0.001))$ & 4.464$\times 10^5$ &  4.694$\times 10^5$ &$4.463 \times 10^5$& 4.463$\times 10^5$ \\
$\sigma(\sum_{i=1}^{6}c_i(0.001))$ &16.226 & 19.444 &6.071&-- \\ \hline\hline
\end{tabular}\label{tab3}
\end{table}

It is important to point out that the issue of stiffness arises when solving the stochastic models for these scenarios. This puts an additional constraint in the probability calculations. As in the previous example, an analysis of the standard deviations in Tables \ref{tab2} and \ref{tab3} indicates that the stochastic effects need to be taken under consideration, since the values obtained for the mean and the standard deviation of the neutron density are of the same order of magnitude.

\begin{figure}[!ht]
\centering
\includegraphics[width=12.0cm,height=9.0cm]{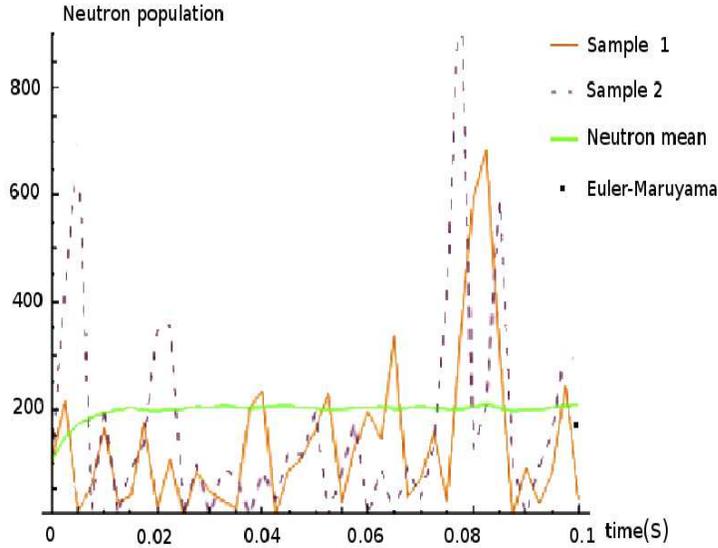} 
\caption{Neutron Density for six Precursor Groups with reactivity $\rho=0.007$.}
\label{fig:wollmann1}
\end{figure}

Besides the evaluation for a fixed time $t=0.1s$ by the Euler-Maruyama approach, we also generate the time line (Figue \ref{fig:wollmann1}) of the neutron density and compare two Monte Carlo realizations (Sample 1 and Sample 2) with the mean value of the neutron density after averaging over a sufficiently large set of samples.

\subsection{Linear Reactivity}\label{sec3.2}

\begin{figure}[!th]
\centering
\includegraphics[width=12.0cm,height=9.0cm]{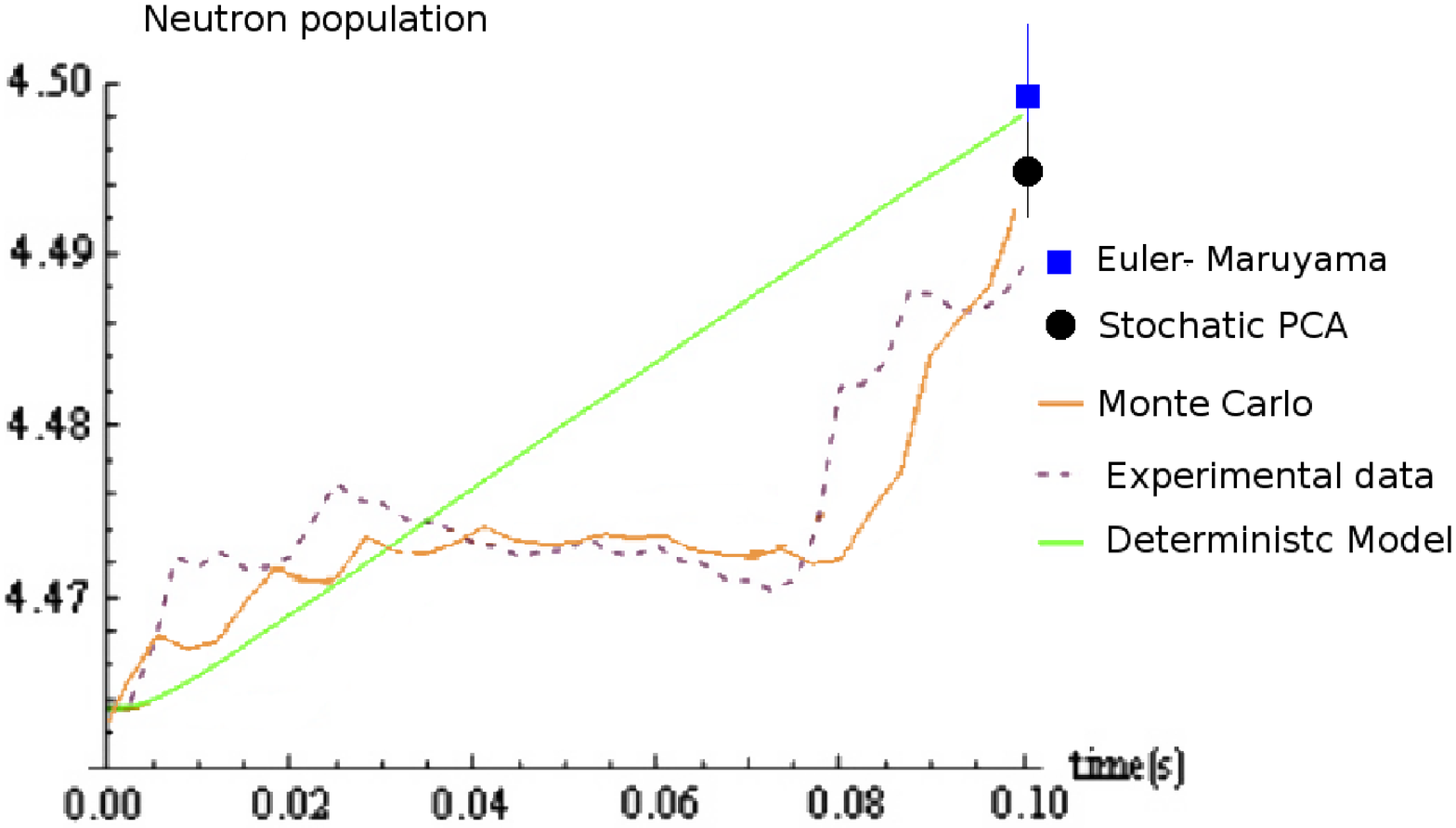} 
\caption{Neutron Density for one precursor group and reactivity $\rho(t)=0.25t$.}
\label{fig:wollmann2}
\end{figure}

The example discussed in this section is, to the best of our knowledge, the first study of this kind that considers time-dependent reactivity. We provide Monte Carlo results for an example with one precursor group and linear reactivity (see Figure \ref{fig:wollmann2}) and compare our findings to experimental data \cite{Han60} as well as to the deterministic model prediction \cite{pevi09,sile14}. For the time $t=0.1s$, the Stochastic PCA and Euler-Maruyama results are indicated. The parameter set used for this simulation is $\lambda_1=0.1$, $\beta_1=0.005$, $\nu=2.5$, $l=0.00001$ with time dependent reactivity $\rho(t)=0.25t$ and with initial condition
$\vec{x}(0)=100[1,\frac{\beta_1}{\lambda_1l}]^T.$

We note that, while the Deterministic Model yields a curve with the correct qualitative behavior, it fails to provide any information on the stochastic fluctuations of the neutron population over time. Clearly, a model that can predict these fluctuations would be an improvement over the deterministic approach.

\section{Discussion}\label{sec4}

From the phenomenological point of view, it is evident that one needs to take under consideration the stochastic effects in order to compute the neutron density. This is confirmed by the results of the simulations we have presented, where we see that the values for the mean and standard deviation of the neutron density can be of the same order of magnitude. The examples presented here also suggest that the fluctuations in the precursor concentrations are small. This behavior arises from the stochastic nature of decay; specifically, from the property of time homogeneity inherent to the radioactive decay law.

The present work is the first one in a sequence, in which reactivity of time dependent scenarios and the effects of stochastic moments are studied. This will be done by solving the stochastic equation in a hierarchic fashion: first, the deterministic part of the problem is solved, and then the solution is modified by including the stochastic moments. This contrasts with the procedures currently found in the literature, which make use of the roots of the inhour equation. One of the main difficulties encountered refers to the stiffness of the problem, which imposes severe restrictions on the calculation of the event probabilities. In a future work these issues will be addressed in an optimised solution procedure. 

\subsection*{Acknowledgments}
M.T.V. and R.V. would like to thank CNPq and M.W.d.S. would like to thank CAPES for financial support.

\end{document}